\documentclass[a4paper, 10pt, oneside, onecolumn, notitlepage, final]{article}
\usepackage[text={160mm, 240mm}, left=25mm, vmarginratio=1:1]{geometry}
\usepackage{graphicx}
\usepackage{amssymb, amsfonts, amsmath, amsthm, mathtools, mathrsfs}
\usepackage{upgreek}
\usepackage[usenames, dvipsnames]{xcolor}
\usepackage[colorlinks, linkcolor=blue, anchorcolor=red, citecolor=blue]{hyperref}
\usepackage[backend=biber, sorting=none, maxbibnames=3, minbibnames=3, style=numeric-comp]{biblatex}
\usepackage{braket} 
\usepackage{authblk} 
\usepackage{orcidlink}
\usepackage{abstract}
\usepackage{titlesec}
\usepackage[T1]{fontenc}
\usepackage[utf8]{inputenc}
\usepackage{lmodern}
\usepackage[english]{babel}

\titleformat{\section}{\large\bfseries\centering}{\thesection}{0.5 em}{}
\titleformat{\subsection}{\normalsize\bfseries\centering}{\thesubsection}{0.5 em}{}

\begin{document}

\title{\Large\bfseries Upper Bounded Current Fluctuations in One-Dimensional Driven Transport Systems}

\author[1]{\normalsize Jiayin Gu\orcidlink{0000-0002-9868-8186}\thanks{\texttt{gujiayin@njnu.edu.cn}}}
\author[2,3]{\normalsize Fan Zhang\orcidlink{0000-0002-7466-6898}\thanks{\texttt{van314159@pku.edu.cn}}}

\affil[1]{\normalsize School of Physics and Technology, Nanjing Normal University, Nanjing 210023, China}
\affil[2]{\normalsize School of Physics, Peking University, Beijing 100871, China}
\affil[2]{\normalsize RIKEN Center for Emergent Matter Science, RIKEN, Saitama, 351-0198, Japan}

\date{}
\maketitle

\begin{abstract}
We conjecture that the current fluctuations in one-dimensional driven transport systems obey an upper bound determined by the mean current and the driving force. This inequality originates from repulsive interactions between transporting particles, and the bound is approached both in near-equilibrium systems and in far-from-equilibrium systems with weak interactions. We first propose a coarse-grained model describing random particle exchanges between two reservoirs with constant rates, from which the upper bound emerges. We then rigorously prove the inequality in quantum ballistic transport systems. Finally, we demonstrate its validity in two specific diffusive systems: the exclusion process, for which the inequality can be proven, and charged-particle transport, for which numerical evidence supports the inequality.
\end{abstract}

\section{Introduction}

Understanding the transport behavior of interacting particles is a central question in nonequilibrium statistical physics~\cite{Spohn_1991, Kipnis_1999, Liggett_1999}. Recent decades have seen progress through extensive studies of stochastic lattice systems, such as the exclusion process~\cite{Harris_JApplProbab_1965, Spitzer_AdvMath_1970}, the zero-range process~\cite{Spitzer_AdvMath_1970, Levine_JStatPhys_2005}, and the contact process~\cite{Harris_AnnProbab_1974}. Moreover, general statements can also be made on systems in far-from-equilibrium regimes. The microscopic reversibility constrains the current fluctuations in nonequilibrium steady-state systems, leading to the establishment of the fluctuation theorem~\cite{Evans_PhysRevLett_1993, Evans_AdvPhys_2002, Gallavotti_PhysRevLett_1995, Gallavotti_JStatPhys_1995, Kurchan_JPhysA_1998, Lebowitz_JStatPhys_1999, Maes_JStatPhys_1999, Harris_JStatMech_2007, Gaspard_NewJPhys_2013}. Close to equilibrium, this theorem reduces to well-known fluctuation-dissipation relations, such as the Green-Kubo formulas~\cite{Zwanzig_AnnRevPhysChem_1965}. They also imply the Onsager reciprocal relations~\cite{Casimir_RevModPhys_1945} as well as the generalized nonlinear ones up to arbitrary orders~\cite{Andrieux_JChemPhys_2004, Gaspard_NewJPhys_2013, Gu_PhysRevE_2019, Barbier_JPhysA_2018, Barbier_JPhysA_2019, Gu_JStatMech_2020, Wu_PhysRevE_2022, Zhang_PhysRevE_2023}. In recent years, great advances were initiated by the discovery of the so-called thermodynamic uncertainty relations (TURs)~\cite{Barato_PhysRevLett_2015, Gingrich_PhysRevLett_2016, Horowitz_NatPhys_2020}, which provide general lower bounds on current fluctuations in terms of its mean value and overall entropy production. These relations are interpreted as the thermodynamic cost of precision and find applications in obtaining upper bounds on the efficiency of molecular motors~\cite{Hwang_JPhysChemLett_2018}, inferring entropy production~\cite{Manikandan_PhysRevLett_2020}, and even understanding the dynamics of sperm~\cite{Maggi_PRXLife_2023}. Notably, the TUR can itself be derived from the fluctuation theorem~\cite{Timpanaro_PhysRevLett_2019}.

\par Despite these advances, the transport behavior for the current in one-dimensional systems remains incompletely understood. For example, how do the interactions between the transporting particles influence the current fluctuations? In this paper, we try to answer this question, and this leads us to discover an upper bound on current fluctuations. For a one-dimensional nonequilibrium driven transport system whose mean current is $J$ and the diffusivity is $D$~\footnote{We denote integrated current over time interval $[0,\,t]$ by $Z$, then the mean current and its diffusivity are defined by $J\equiv\lim_{t\to\infty}\langle Z_t\rangle/t$ and $D\equiv\lim_{t\to\infty}\langle(Z-Jt)^2\rangle/(2t)$, where $\langle \rangle$ stands for the statistical average.}, we conjecture an inequality stating that the Fano factor $2D/J$ is upper bounded,
\begin{align}
\frac{2D}{J}\le \coth(A/2) \text{,} \label{eq_inequality}
\end{align}
where $A$ stands for the affinity driving the system out of equilibrium~\cite{DeDonder_1936, Prigogine_1967}. For an isothermal system at inverse temperature $\beta$, the affinity is defined in terms of the difference of chemical potentials between the left and right reservoirs, $A=\beta(\mu_{\rm L}-\mu_{\rm R})$. The upper bound is saturated when the system is close to equilibrium. This follows from the fluctuation-dissipation relation $J=DA$, which ensures that the inequality~(\ref{eq_inequality}) reduces to it correctly in the limit $A\to 0$. In addition, the original TUR~\cite[Eq.~(4) in][]{Barato_PhysRevLett_2015} can be reformulated as $D\ge J/A$, where the implied lower bound tends to coincide with our conjectured upper bound as $A\to 0$. Moreover, when the system is driven far from equilibrium, the upper bound is also saturated in the case where the interactions between the transporting particles are absent. The suppressed current fluctuations can be understood intuitively by recalling the subdiffusion of tagged particles in a medium~\cite{Arratia_AnnProbab_1983, Spohn_JStatPhys_1990, Imamura_PhysRevLett_2017}. In the exclusion process, where the tracers cannot bypass neighboring particles and consequently undergo subdiffusion, the growth of mean square displacement (MSD) scales with the square root of time. This is much slower compared with the normal diffusion case, where the MSD of a tracer grows linearly with time.

\par Recently, we noted a related work~\cite{BakewellSmith_PhysRevLett_2023}, in which an upper bound is proved on the fluctuations of flux observables of trajectories. This bound relies on an input mathematical quantity that is not operationally accessible. In contrast, our upper bound is more physical. An interesting question is whether, in the case where our bound is saturated by current fluctuations, their bound is lower bounded by our bound? The answer to this question is promising for future explorations.

\par The inequality~(\ref{eq_inequality}) is expected to hold in different systems. In the following, we demonstrate how the inequality~(\ref{eq_inequality}) is derived. First, a coarse-grained model describing the long-time transport behavior is proposed. From this, the upper bound is obtained. Second, the inequality is rigorously proven for quantum ballistic transport. Then, the validity of the inequality is demonstrated in two specific driven diffusive systems. In addition, the inequality has also been numerically shown to hold in diffusion-reaction systems; the details are presented elsewhere~\cite{Gu_arXiv_2024}.

\section{Coarse-Grained Model}

We start by proposing a coarse-grained model at the highest level of description for one-dimensional driven transport systems. In this model, particles are randomly exchanged between two reservoirs at constant rates $W_+$ and $W_-$. For this model, the mean current, its diffusivity as well as the affinity given by $J=W_+-W_-$, $D=(W_++W_-)/2$, and $A=\ln(W_+/W_-)$, respectively~\cite{SupplementalMaterial_misc}. This yields the exact relation:
\begin{align}
\frac{2D}{J}=\coth(A/2) \text{.} \label{eq_equality}
\end{align}
The immediate implication is that if this coarse-grained model accurately describes the long-time transport behavior in general nonequilibrium systems, then this equality can be employed to test the fluctuation theorem indirectly. Since the fluctuation theorem is the property of large deviations~\cite{Lebowitz_JStatPhys_1999, Ellis_2006, Touchette_PhysRep_2009, Dembo_2010}, it is generally very difficult to access rare events that are essential to the direct test of the fluctuation theorem. However, it is relatively easy to measure the mean current $J$ and its diffusivity $D$. This equality has previously been used to test the fluctuation theorem for the charge transport in diodes~\cite{Gu_PhysRevE_2018}. In the following, it will be shown that, for general nonequilibrium systems where interactions between transporting particles are present, this equality is violated, replaced by the inequality~(\ref{eq_inequality}).

\section{Quantum Ballistic Transport}

We now consider a one-dimensional nonequilibrium system for fermionic ballistic transport. The setup can be an atomic chain, a nanowire, or a series of quantum dots that are connected to two reservoirs at the uniform inverse temperature $\beta$ and imbalanced chemical potentials, $\mu_{\rm L}>\mu_{\rm R}$. The mean current and its diffusivity are given by the Landauer-Buttike formula~\cite{Nazarov_2009},
\begin{align}
& J=\int\frac{{\rm d}\varepsilon}{2\pi}\mathbb{T}(\varepsilon)\left(n_{\rm L}-n_{\rm R}\right) \text{,} \\
& D=\int\frac{{\rm d}\varepsilon}{4\pi}\mathbb{T}(\varepsilon)\left[(n_{\rm L}\bar{n}_{\rm R}+\bar{n}_{\rm L}n_{\rm R})-(n_{\rm L}-n_{\rm R})^{2}\right] \text{,} 
\end{align}
where the transmission function $\mathbb{T}(\varepsilon)\in[0,1]$ is determined by the microscopic model of the system and calculated by the nonequilibrium Green's function~\cite{Kamenev_2011}. The Fermi distribution of electrons with energy $\varepsilon$ of the reservoir $\alpha={\rm L},{\rm R}$ is $n_{\alpha}(\varepsilon)=1/\left[{\rm e}^{\beta(\varepsilon-\mu_{\alpha})}+1\right]$ and $\bar{n}_{\alpha}(\varepsilon)=1-n_{\alpha}(\varepsilon)$ denotes the distribution of unoccupied states. Then, the Fano factor can be expressed as
\begin{align}
\frac{2D}{J}=\frac{\int\frac{{\rm d}\varepsilon}{2\pi}\mathbb{T}(\varepsilon)(n_{\rm L}\bar{n}_{\rm R}+\bar{n}_{\rm L}n_{\rm R})}{\int\frac{{\rm d}\varepsilon}{2\pi}\mathbb{T}(\varepsilon)(n_{\rm L}\bar{n}_{\rm R}-\bar{n}_{\rm L}n_{\rm R})}-\frac{\int\frac{{\rm d}\varepsilon}{2\pi}\mathbb{T}(\varepsilon)(n_{\rm L}-n_{\rm R})^{2}}{\int\frac{{\rm d}\varepsilon}{2\pi}\mathbb{T}(\varepsilon)(n_{\rm L}\bar{n}_{\rm R}-\bar{n}_{\rm L}n_{\rm R})} \leq\frac{\int\frac{{\rm d}\varepsilon}{2\pi}\mathbb{T}(\varepsilon)(n_{\rm L}\bar{n}_{\rm R}+\bar{n}_{\rm L}n_{\rm R})}{\int\frac{{\rm d}\varepsilon}{2\pi}\mathbb{T}(\varepsilon)(n_{\rm L}\bar{n}_{\rm R}-\bar{n}_{\rm L}n_{\rm R})} \text{,} \label{eq_ineq1}
\end{align}
where $n_{\rm L}\bar{n}_{\rm R}\ge\bar{n}_{\rm L}n_{\rm R}$ was assumed. The last term in Eq.~(\ref{eq_ineq1}) can be simplified,
\begin{align}
{\rm last\,term}=\frac{\int\frac{{\rm d}\varepsilon}{2\pi}\mathbb{T}(\varepsilon)\bar{n}_{\rm L}n_{\rm R}({\rm e}^{A}+1)}{\int\frac{{\rm d}\varepsilon}{2\pi}\mathbb{T}(\varepsilon)\bar{n}_{\rm L}n_{\rm R}({\rm e}^{A}-1)}=\coth(A/2) \text{,}
\end{align}
where the relation for the affinity 
\begin{align}
A=\ln\frac{n_{\rm L}\bar{n}_{\rm R}}{\bar{n}_{\rm L}n_{\rm R}}=\beta\left(\mu_{\rm L}-\mu_{\rm R}\right) \label{eq_A1}
\end{align}
was used. Thus, the inequality~(\ref{eq_inequality}) is rigorously proved. We explain that it is the interaction due to the Pauli exclusion principle that suppresses the current fluctuations. We also point out that the transmission function encoding the detailed dynamics is irrelevant in the above proof, manifesting that the proof applies to a broad class of different quantum systems. This also contrasts sharply with the original version of TUR that can be violated in quantum systems~\cite{Agarwalla_PhysRevB_2018}.

\begin{figure}
\centering
\begin{minipage}[t]{0.6\hsize}
\resizebox{1.0\hsize}{!}{\includegraphics{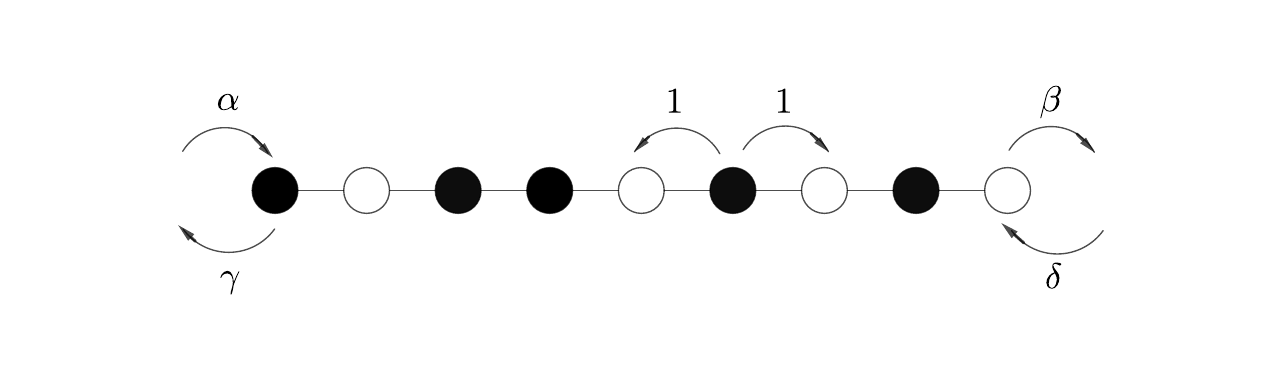}}
\end{minipage}
\begin{minipage}[t]{0.6\hsize}
\resizebox{1.0\hsize}{!}{\includegraphics{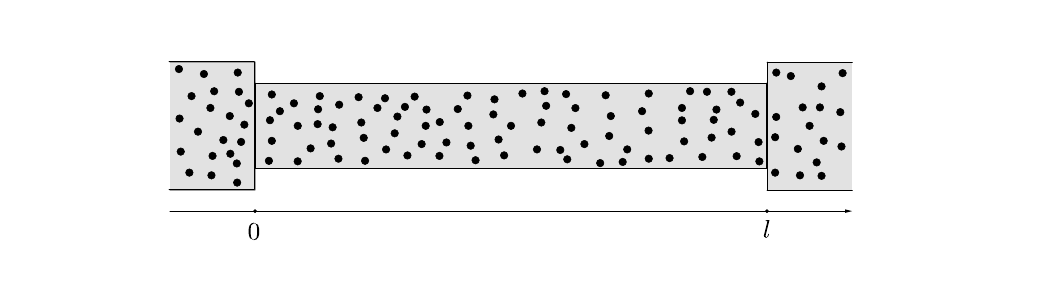}}
\end{minipage}
\caption{Schematic representations of driven diffusive systems. Top panel: The symmetric simple exclusion process. The black (white) circles denote occupied (vacant) sites. Bottom panel: Charged particles transporting in a conductive channel.}
\label{fig_systems}
\end{figure}

\section{Symmetric Simple Exclusion Process}

We now turn to a paradigmatic model in nonequilibrium statistical physics, the symmetric simple exclusion process (SSEP)~\cite{Harris_JApplProbab_1965, Spitzer_AdvMath_1970, Derrida_JStatMech_2007, Derrida_JStatPhys_2004, Derrida_JStatMech_2011, Mallick_PhysicaA_2015}. This process describes the hopping of hard-core particles in a one-dimensional chain, driven solely at two boundaries by the inflow and outflow of particles. As shown in the top panel of Figure~\ref{fig_systems}, the particles stochastically hop to a vacant neighboring site at a rate of $1$ in the bulk and enter (exit) at the left and right boundaries at rates $\alpha$ ($\gamma$) and $\delta$ ($\beta$). This system was extensively investigated during the last few decades, and therefore a wealth of analytical knowledge has been accumulated~\cite{Krapivsky_2010, Katz_JStatPhys_1983, Lebowitz_JStatPhys_1988, Derrida_JStatMech_2007, Derrida_JStatMech_2011}. For this system, the densities of the left and right reservoirs are identified as $\rho_{\rm L}=\alpha/(\alpha+\gamma)$ and $\rho_{\rm R}=\delta/(\beta+\delta)$, respectively. In addition, two quantities $a=1/(\alpha+\gamma)$ and $b=1/(\beta+\delta)$ are introduced for later use. If these two densities are unequal, the system relaxes to a nonequilibrium steady state after some transient time. In this case, the mean current is given by
\begin{align}
J=(\rho_{\rm L}-\rho_{\rm R})/\tilde{L} \text{,} \label{eq_SSEP_J}
\end{align}
where $\tilde{L}\equiv L+a+b-1$ and $L$ denotes the number of sites. Moreover, the diffusivity of the current can also be given by~\cite{Derrida_JStatPhys_2004, Derrida_JStatMech_2007, Derrida_JStatMech_2011}
\begin{align}
D= \frac{1}{2\tilde{L}}\left(\rho_{\rm L}+\rho_{\rm R}-2\rho_{\rm L}\rho_{\rm R}\right)+\frac{a(a-1)(2a-1)+b(b-1)(2b-1)-\tilde{L}(\tilde{L}-1)(2\tilde{L}-1)}{6\tilde{L}^3(\tilde{L}-1)}(\rho_{\rm L}-\rho_{\rm R})^2 \text{,} \label{eq_SSEP_D}
\end{align}
Analogous to Eq.~(\ref{eq_A1}) for quantum ballistic transport, the affinity can be evaluated~\cite{Derrida_JStatPhys_2004, Derrida_JStatMech_2007, Derrida_JStatMech_2011},
\begin{align}
A=\ln \frac{\rho_{\rm L}\bar{\rho}_{\rm R}}{\bar{\rho}_{\rm L}\rho_{\rm R}}=\ln\frac{\alpha\beta}{\gamma\delta} \label{eq_SSEP_A} \text{,}
\end{align}
where $\bar{\rho}_{\rm L}\equiv 1-\rho_{\rm L}$ and $\bar{\rho}_{\rm R}\equiv 1-\rho_{\rm R}$ are densities quantifying vacancy. This affinity can be identified through Schnakenberg's graph analysis of the master equation for the jump process~\cite{Schnakenberg_RevModPhys_1976}. It is given as the ratio of the product of the transition rates along the cyclic path over the product of the transition rates along the reversed cyclic path~\cite{SupplementalMaterial_misc}. The implication of affinities identified from cycles for the graph leads to the establishment of the fluctuation theorem in the early form~\cite{Jiang_2004}.

\par
\begin{figure}
\centering
\begin{minipage}[t]{0.45\hsize}
\resizebox{1.0\hsize}{!}{\includegraphics{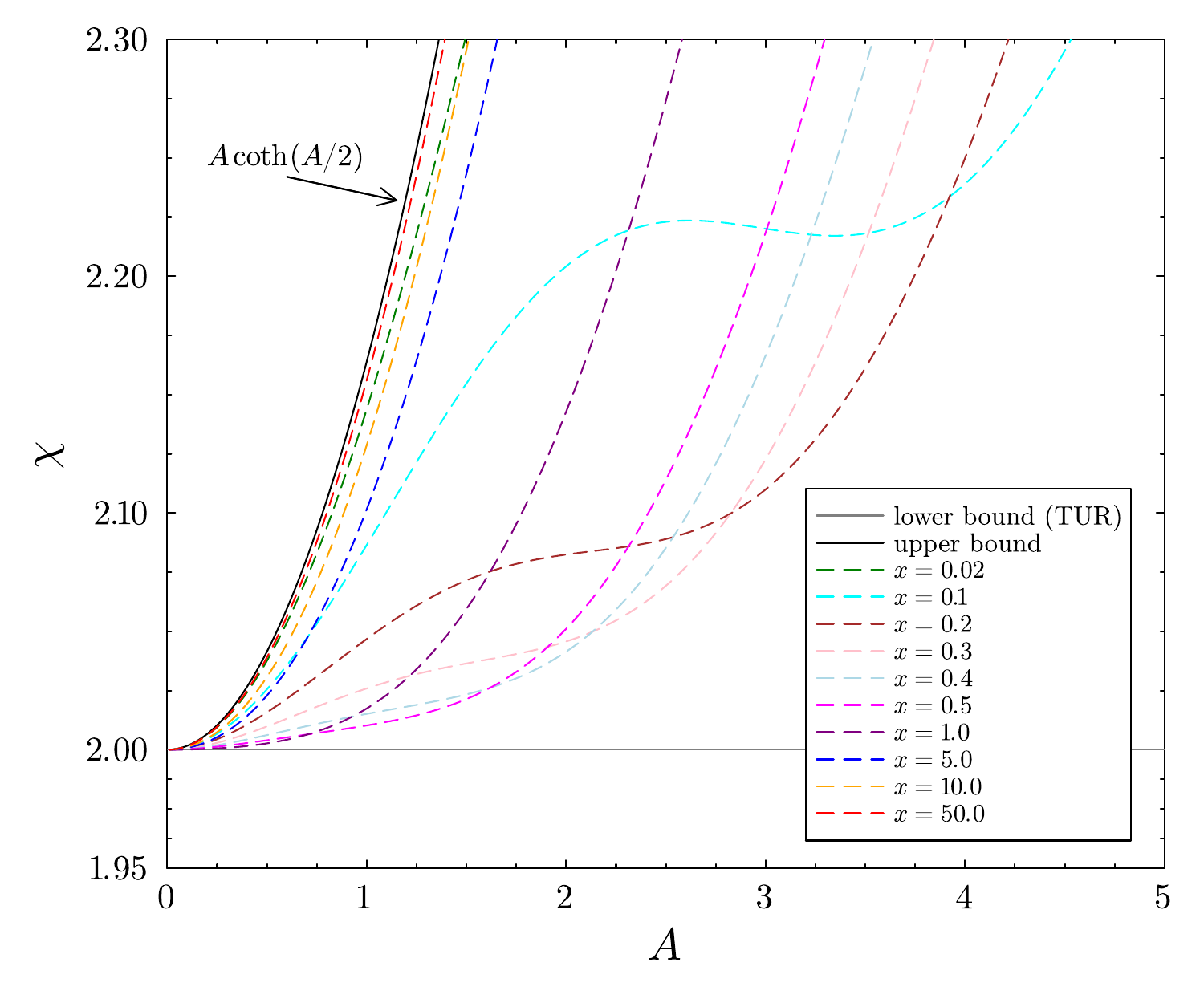}}
\end{minipage}
\begin{minipage}[t]{0.45\hsize}
\resizebox{1.0\hsize}{!}{\includegraphics{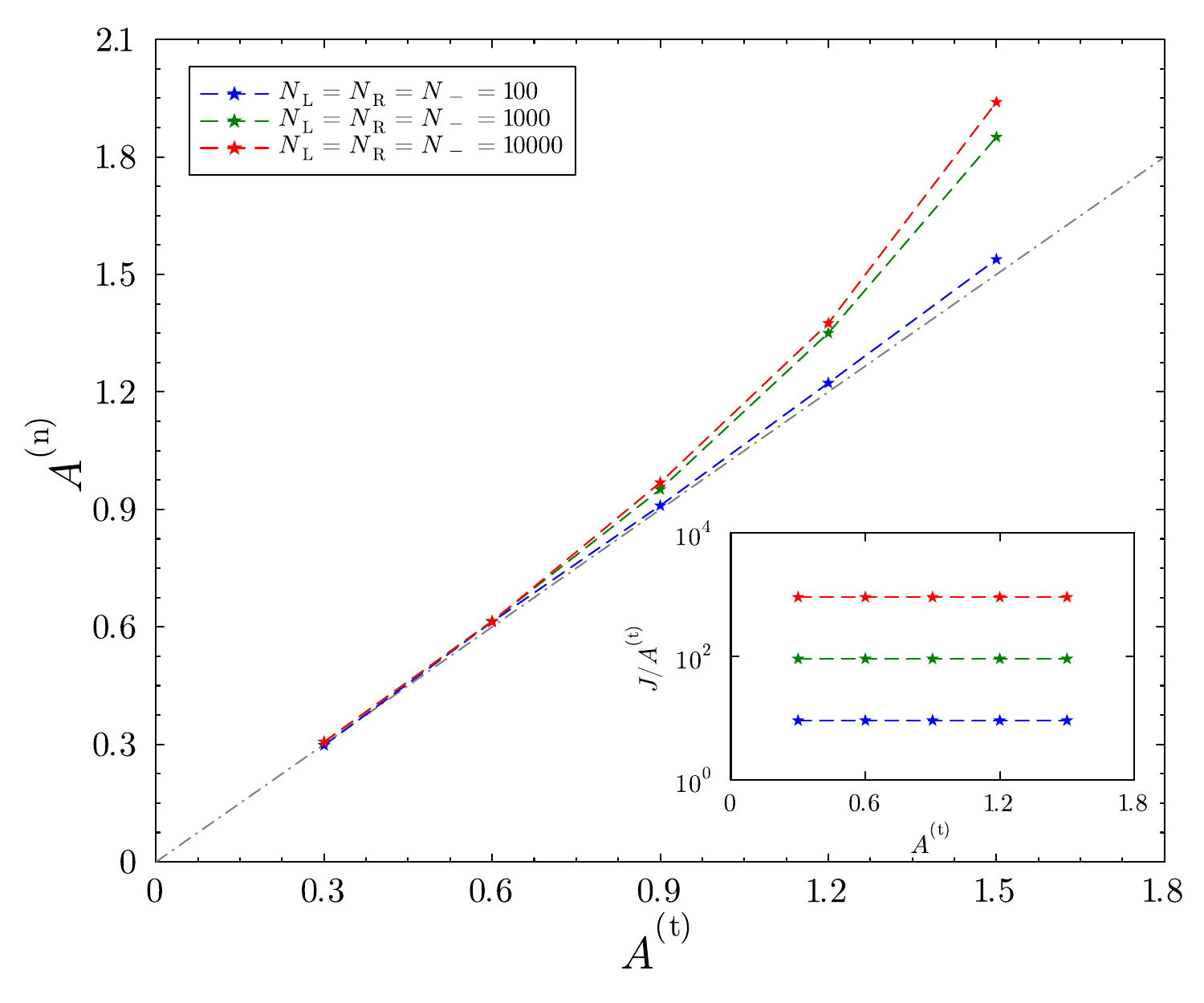}}
\end{minipage}
\caption{(Color online) Graphical representations of the bounded current fluctuations. Left panel: The behavior of $\chi\equiv 2DA/J$ as a function of $A$ with different particle densities (controlled by a parameter $x\equiv \gamma/\alpha$ with a small value for high density and a large value for low density) for SSEP. The parameter values are $\gamma=\beta=1$ and $L=10$. Moreover, the condition $y={\rm e}^Ax$ with $y\equiv\beta/\delta$ is also imposed. The black solid line is the affinity-dependent upper bound $A\coth(A/2)$, and the gray solid line is the lower bound $2$ from the TUR. The dashed lines are depicted from Eqs.~(\ref{eq_SSEP_J})-(\ref{eq_SSEP_A}). Right panel: The comparison between numerical affinities~(\ref{eq_An}) and theoretical affinities~(\ref{eq_At}) for the transport of charged particles. It is set that $\bar{N}_{\rm L}=\bar{N}_{\rm R}=N_-=n_-\Omega$, and takes different values in three cases, as shown in the legend. The asterisks are numerical points with dashed lines joining them. The dot-dash line indicates the equality between both kinds of affinities. The parameter values used in simulation are $\beta=e=1.0$, $D=\epsilon=0.01$, $\Omega=10000$, $\Delta x=0.1$, $L=10$. The numerical affinities are computed with the mean current $J$ and its diffusivity $D$ evaluated over time interval $[0,\,5000]$ with $20000$ data.}
\label{fig_results}
\end{figure}

 At this point, we are ready to calculate the Fano factor with the expressions~(\ref{eq_SSEP_J}) and~(\ref{eq_SSEP_D}), reading
\begin{align}
\frac{2D}{J} & = \frac{\rho_{\rm L}+\rho_{\rm R}-2\rho_{\rm L}\rho_{\rm R}}{\rho_{\rm L}-\rho_{\rm R}}+\frac{a(a-1)(2a-1)+b(b-1)(2b-1)-\tilde{L}(\tilde{L}-1)(2\tilde{L}-1)}{3\tilde{L}^2(\tilde{L}-1)}(\rho_{\rm L}-\rho_{\rm R}) \text{.} \label{eq_ineq2}
\end{align}
Remarkably, the first term on the right-hand side in Eq.~(\ref{eq_ineq2}) can be further simplified as
\begin{align}
\frac{\rho_{\rm L}+\rho_{\rm R}-2\rho_{\rm L}\rho_{\rm R}}{\rho_{\rm L}-\rho_{\rm R}} & = \frac{\rho_{\rm L}\bar{\rho}_{\rm R}+\rho_{\rm R}\bar{\rho}_{\rm L}}{\rho_{\rm L}\bar{\rho}_{\rm R}-\rho_{\rm R}\bar{\rho}_{\rm L}} = \frac{\alpha\beta+\delta\gamma}{\alpha\beta-\delta\gamma} \nonumber \\
& = \frac{{\rm e}^A+1}{{\rm e}^A-1}= \coth(A/2) \text{,}
\end{align}
which yields the expected result. To prove that the inequality~(\ref{eq_inequality}) holds for this system, we only need to show that the second term is nonpositive. Taking $L=1$, we have
\begin{align}
{\rm second\;term}=-\frac{2ab}{(a+b)^2}\left(\rho_{\rm L}-\rho_{\rm R}\right) \text{,}
\end{align}
which is non-positive~\cite{SupplementalMaterial_misc}. Here, $\rho_{\rm L}\ge\rho_{\rm R}$ was implicitly assumed. When $L\ge 2$, the denominator is always positive, and the numerator decreases as $L$ increases. The maximum of the numerator at $L=2$ is
\begin{align}
-6\left(a^2b+a^2+ab+ab^2+b^2\right)\left(\rho_{\rm L}-\rho_{\rm R}\right) \text{,}
\end{align}
indicating that the second term is also non-positive, as expected~\cite{SupplementalMaterial_misc}. The equality sign in Eq.~(\ref{eq_inequality}) for this system asymptotically holds when $1\gg\rho_{\rm L}>\rho_{\rm R}>0$, i.e., the two densities are both very low. This translates to the situation where the exclusion between diffusive particles is negligible. The graphical presentations are shown in the left panel of Figure~\ref{fig_results}, where the defined quantity $\chi\equiv 2DA/J$ is bounded from below by $2$ (TUR) and bounded from above by $A\coth(A/2)$. It is clear that the upper bound is saturated in the low-density limit. Interestingly, the upper bound also appears to be saturated in the high-density limit.

\section{Transport of Charged Particles}

Next, we turn to the transport of charged particles in a conductive channel at the mesoscopic level. As shown in the bottom panel of Figure~\ref{fig_systems}, the channel can be thought of as a rod extending along the $x$-axis from $0$ to $l$. In the transverse $y$- and $z$- directions, the section area is denoted by $\Sigma$. Two types of charged particles are supposed to be distributed in this channel: mobile positive-charged particles with the density $n(x)$, and anchored negative-charged particles with uniform density $n_-(x)=n_-$. The charge density is therefore given by $\rho(x)=e[n(x)-n_-]$, where $e=|e|$ represents the elementary electric charge. Both terminals are in contact with reservoirs that fix the densities of mobile positive-charged particles with $n_{\rm L}$, $n_{\rm R}$ and electric potentials with $\Phi_{\rm L}$, $\Phi_{\rm R}$. The electric potential $\Phi(x)$ across the channel fluctuates and is determined through the Poisson equation by the fluctuating charge distributions as well as the boundary conditions.

\par The basic idea for simulating the system is to first discretize the system in space and then establish the master equation to describe the stochastic evolution of the system state~\cite{Andrieux_JStatMech_2009}. For this purpose, the channel is discretized into cells, each with width $\Delta x$ and volume $\Omega=\Sigma\Delta x$. There are a total of $L=l/\Delta x$ such cells. The reservoirs are modeled as two cells that contain a fixed number of positive-charged particles $\bar{N}_{\rm L}=n_{\rm L}\Omega$, $\bar{N}_{\rm R}=n_{\rm R}\Omega$. The system state is specified by the numbers $\{N_1,N_2,\cdots,N_L\}$ of the mobile charged particles in the intermediate cells. It evolves in time according to the network
\begin{equation}
\resizebox{0.85\hsize}{!}{$
\begin{array}{ccccccccccccccccc}
\bar{N}_{\rm L} & \xrightleftharpoons[W_0^{(-)}]{W_0^{(+)}} & N_1 & \xrightleftharpoons[W_1^{(-)}]{W_1^{(+)}} & N_2 & \xrightleftharpoons[W_2^{(-)}]{W_2^{(+)}} & \cdots & \xrightleftharpoons[W_{i-1}^{(-)}]{W_{i-1}^{(+)}} & N_i & \xrightleftharpoons[W_i^{(-)}]{W_i^{(+)}} & \cdots & \xrightleftharpoons[W_{L-2}^{(-)}]{W_{L-2}^{(+)}} & N_{L-1}& \xrightleftharpoons[W_{L-1}^{(-)}]{W_{L-1}^{(+)}} & N_L & \xrightleftharpoons[W_L^{(-)}]{W_L^{(+)}} & \bar{N}_{\rm R} \text{,} \label{eq_network}
\end{array}$}
\end{equation}
with the transition rates given by
\begin{align}
& W_i^{(+)}=\frac{D}{\Delta x^2}\psi(\Delta U_{i,i+1})N_i \text{,} \\
& W_i^{(-)}=\frac{D}{\Delta x^2}\psi(\Delta U_{i+1,i})N_{i+1} \text{.}
\end{align}
Here, $D$ denotes the diffusion coefficient of mobile particles, not to be confused with the current diffusivity. $\Delta U_{i,i+1}$ is the intrinsic energy change of the system associated with one particle jumping from the $i$-th to $(i+1)$-th cell and is given by~\cite{Andrieux_JStatMech_2009}
\begin{align}
\Delta U_{i,i+1}=e(\Phi_{i+1}-\Phi_i)+\frac{e^2L\Delta x^2}{2(L+1)\epsilon\Omega} \text{,}
\end{align}
where $\epsilon$ is the permittivity. The function $\psi(\Delta U)$ is defined as
\begin{align}
\psi(\Delta U)=\frac{\beta\Delta U}{\exp(\beta\Delta U)-1} \text{,}
\end{align}
which guarantees the detailed balance in equilibrium, $\psi(\Delta U)=\psi(-\Delta U)\exp(-\beta\Delta U)$. The master equation can be readily written down from the network~(\ref{eq_network}) and random trajectories can be generated with the Gillespie algorithm~\cite{Gillespie_JComputPhys_1976}. In numerical simulations, a much faster algorithm based on a Langevin-type stochastic process is implemented~\cite{SupplementalMaterial_misc}.

In numerical simulations, we impose the boundary condition $\bar{N}_{\rm L}=\bar{N}_{\rm R}=\bar{N}$ so that the affinity in theory is given by the potential difference between two reservoirs,
\begin{align}
A^{({\rm t})}=\beta e(\Phi_{\rm L}-\Phi_{\rm R}) \text{.} \label{eq_At}
\end{align}
The integrated current crossing a section over a large time interval is counted. The mean current $J$ and its diffusivity $D$ can be extracted from the counting statistics. If the long-time behavior of the transport was accurately described by the coarse-grained model, then we have $W_+=D+J/2$ and $W_-=D-J/2$. Consequently, we can estimate the affinity numerically from these two cumulants 
\begin{align}
A^{({\rm n})}=\ln\frac{W_+}{W_-}=\ln\frac{2D+J}{2D-J} \text{.} \label{eq_An}
\end{align}
The results are displayed in the right panel of Figure~\ref{fig_results}, where we compared $A^{({\rm n})}$  and $A^{({\rm t})}$ in three cases of different particle densities. We notice that $A^{({\rm n})}$ does not always agree with $A^{({\rm t})}$, the former is generally greater than or equal to the latter, $A^{({\rm n})}\ge A^{({\rm t})}$, leading to the inequality~(\ref{eq_inequality}). When the system is near equilibrium, they take small values and are approximately equal to each other. However, when the system is driven far from equilibrium, the approximate equality is only found in the case of very low particle densities. It can be argued that the charged particles with low densities play a negligible role in determining the potential $\Phi(x)$ by the Poisson equation, the fluctuating electric field in this case can be approximated as a static background field. It can also be exactly proven that, in the case of the low-density limit, the long-time behavior of the transport can indeed be accurately described by the coarse-grained model, with the equivalent global transition rates given by~\cite{SupplementalMaterial_misc}
\begin{align}
& W_+=\frac{D\bar{N}}{\Delta x^2(L+1)}\frac{\beta e(\Phi_{\rm R}-\Phi_{\rm L})}{\exp\left[\beta e(\Phi_{\rm R}-\Phi_{\rm L})\right]-1} \text{,} \\
& W_-=\frac{D\bar{N}}{\Delta x^2(L+1)}\frac{\beta e(\Phi_{\rm L}-\Phi_{\rm R})}{\exp\left[\beta e(\Phi_{\rm L}-\Phi_{\rm R})\right]-1} \text{,}
\end{align}
where the $D$ here is the diffusion coefficient of mobile particles. In this case, the affinity~(\ref{eq_At}) can be recovered from Eq.~(\ref{eq_An}), and the equality sign in the inequality~(\ref{eq_inequality}) holds. This is explained that the electrostatic repulsive interactions between charged particles play no role, and the transport of charged particles is reduced to normal driven Brownian motion. In contrast, when electrostatic repulsive interactions are present, current fluctuations are suppressed. In addition, the inset of the right panel of Figure~\ref{fig_results} shows that the mean current follows the Ohm law, not affected by the repulsive interactions,

\section{Conclusion}

\par In this work, we have conjectured an inequality that sets an upper bound for current fluctuations in one-dimensional driven transport systems. The proposed coarse-grained model is simple yet illuminating; it enables quantification of the current fluctuations in terms of the mean current and the affinity, which are relatively easy to measure. The inequality is rigorously proven in quantum ballistic transport systems. The validity of the inequality is also supported by two other specific diffusive systems. Although derived through case studies, the inequality captures the essential physics of driven transport of interacting particles. Moreover, the illustrative examples are sufficient to represent a large class of transport systems. To conclude, the inequality represents an important result in nonequilibrium statistical physics. In prospect, for example, it can be used to probe the non-Markovian nature of the long-time transport behavior in systems whose detailed dynamics is modeled Markovian.

\section{Acknowledgment}
Fan Zhang acknowledges H. T. Quan for his encouragement and support. This work was supported by the National Natural Science Foundation of China (NSFC) under the Grant No. 12505048 and the JST Moonshot R\&D under the Grant No. JPMJMS226B.

\appendix

\section{Coarse-Grained Model}

The proposed coarse-grained model describes the long-time transport behavior of one-dimensional nonequilibrium systems. In this model, particles exchange randomly between two reservoirs at constant rates $W_+$ (forward, left to right) and $W_-$ (backward, right to left). Let $Z$ denote the cumulative particle transfers from the left to the right reservoir during the time interval $[0,\,t]$ and ${\cal P}(Z,t)$ be its probability distribution, then the cumulant generating function in terms of the counting parameter $\lambda$ can be defined as
\begin{align}
Q(\lambda)=\lim_{t\to\infty}-\frac{1}{t}\ln\sum_{Z=-\infty}^{+\infty}{\rm e}^{-\lambda Z}{\cal P}(Z,t) \text{.}
\end{align}
The probability distribution obeys the master equation
\begin{align}
\frac{{\rm d}{\cal P}(Z,t)}{{\rm d}t}=\left[W_+\left({\rm e}^{-\partial_Z}-1\right)+W_-\left({\rm e}^{+\partial_Z}-1\right)\right]{\cal P}(Z,t)
\end{align}
We now define the moment generating function of signed cumulated flux,
\begin{align}
G(s,t)\equiv\sum_{Z=-\infty}^{+\infty}s^Z{\cal P}(Z,t) \text{,}
\end{align}
whose time derivative is as follows,
\begin{align}
\partial_tG(s,t) & =W_+\sum_{Z=-\infty}^{+\infty}s^Z{\cal P}(Z-1,t)+W_-\sum_{Z=-\infty}^{+\infty}s^Z{\cal P}(Z+1,t)-(W_++W_-)\sum_{Z=-\infty}^{+\infty}s^Z{\cal P}(Z,t) \nonumber \\
& =\left(W_+s+\frac{W_-}{s}-W_+-W_-\right)G(s,t) \text{.}
\end{align}
For convenience, the initial probability distribution is taken to be ${\cal P}(Z,0)=\delta_{Z,0}$, so
\begin{align}
G(s,0)=\sum_{Z=-\infty}^{+\infty}s^Z\delta_{Z,0}=1 \text{,}
\end{align}
and
\begin{align}
G(s,t)=\exp\left[\left(W_+s+\frac{W_-}{s}-W_+-W_-\right)t\right] \text{.}
\end{align}
The cumulant generating function can thus be obtained,
\begin{align}
Q(\lambda)\equiv\lim_{t\to\infty}-\frac{1}{t}\ln G\left({\rm e}^{-\lambda},t\right)=W_+\left(1-{\rm e}^{-\lambda}\right)+W_-\left(1-{\rm e}^{+\lambda}\right) \text{.} \label{eq_Q1}
\end{align}
It can be easily checked that the cumulant generating function satisfies the Gallavotti-Cohen symmetry
\begin{align}
Q(\lambda)=Q(A-\lambda) \text{,}
\end{align}
where the affinity is defined by
\begin{align}
A=\ln\frac{W_+}{W_-} \text{.}
\end{align}

\section{Symmetric Simple Exclusion Process}

\subsection{Affinity}

In his network theory of Markov jump processes, a graph $G$ can be associated with the master equation such that each state of the system corresponds to a vertex and the different allowed transitions $\omega\rightleftharpoons\omega'$ between the states are represented by edges. In the so-constructed graph, cyclic paths are sequences of edges joining a finite set of vertices and coming back to the starting vertex. Let's denote by $\{\omega\}$ the vertices and $\{e\}$ the edges of the graph, then the affinity of the cyclic path $\cal C$ is defined as
\begin{align}
A({\cal C}) \equiv \ln \prod_{e\in{\cal C}} \frac{W(\omega\stackrel{e}{\rightarrow}\omega')}{W(\omega\stackrel{e}{\leftarrow}\omega')} \label{eq_A_C}
\end{align}
in terms of the ratio of transition rates along the path divided by the transition rates along the reversed path. This affinity characterizes the nonequilibrium constraints imposed by boundaries on the cyclic path. Although the transition rates in Eq.~(\ref{eq_A_C}) normally depend on the mesoscopic states, the so-obtained affinity only depends on the macroscopic thermodynamic force which is of physical importance. For the symmetric simple exclusion process, the cyclic path can be constructed as a particle entering the empty chain from the left, then jumping to the right and eventually exiting the chain. Suppose that the system has $L$ sites. For forward path along the cycle, we have
\begin{align}
\prod_{i=0}^L W_{i,i+1}=\alpha\cdot\underbrace{1\cdot 1\cdots 1}_{L-1}\cdot\beta=\alpha\beta \text{.}
\end{align}
For the reversed path along the cycle, we have
\begin{align}
\prod_{i=0}^L W_{i+1,i}=\gamma\cdot\underbrace{1\cdot 1\cdots 1}_{L-1}\cdot\delta=\gamma\delta \text{.}
\end{align}
Here $W_{i,i+1}$ denotes the transition rate from $i$-th site to $(i+1)$-site. the index $0$ represents the left reservoir and the index $L+1$ represents the right reservoir. Then, the affinity is given by
\begin{align} 
A=\ln\frac{\alpha\beta}{\gamma\delta} \text{.}
\end{align}

\subsection{Inequality}
From the analytical expression of $J$ and $D$, we are readily evaluate the Fano factor $2D/J$, reading
\begin{align}
\frac{2D}{J} & = \frac{\rho_{\rm L}+\rho_{\rm R}-2\rho_{\rm L}\rho_{\rm R}}{\rho_{\rm L}-\rho_{\rm R}}+\frac{a(a-1)(2a-1)+b(b-1)(2b-1)-\tilde{L}(\tilde{L}-1)(2\tilde{L}-1)}{3\tilde{L}^2(\tilde{L}-1)}(\rho_{\rm L}-\rho_{\rm R}) \nonumber \\
& = \frac{\rho_{\rm L}\bar{\rho}_{\rm R}+\rho_{\rm R}\bar{\rho}_{\rm L}}{\rho_{\rm L}\bar{\rho}_{\rm R}-\rho_{\rm R}\bar{\rho}_{\rm L}} +\frac{a(a-1)(2a-1)+b(b-1)(2b-1)-\tilde{L}(\tilde{L}-1)(2\tilde{L}-1)}{3\tilde{L}^2(\tilde{L}-1)}(\rho_{\rm L}-\rho_{\rm R}) \nonumber \\
& = \coth(A/2) +\frac{a(a-1)(2a-1)+b(b-1)(2b-1)-\tilde{L}(\tilde{L}-1)(2\tilde{L}-1)}{3\tilde{L}^2(\tilde{L}-1)}(\rho_{\rm L}-\rho_{\rm R}) \text{,}
\end{align}
where $\tilde{L}=L+a+b-1$. The second term can be proved to be non-positive. Here, we implicitly assumed that $\rho_{\rm L}-\rho_{\rm R}\ge 0$. So $\rho_{\rm L}-\rho_{\rm R}$ is irrelevant in determining the sign of the second term. When $L=1$, the numerator of the second term ($\rho_{\rm L}-\rho_{\rm R}$ ignored) is
\begin{align}
{\rm numerator} & = a(a-1)(2a-1)+b(b-1)(2b-1)-\tilde{L}(\tilde{L}-1)(2\tilde{L}-1) \nonumber \\
  & =  a(a-1)(2a-1)+b(b-1)(2b-1)-(a+b)(a+b-1)(2a+2b-1) \nonumber \\ 
  & =  a(2a^2-3a+1)+b(2b^2-3b+1)-(a+b)(a+b-1)(2a+2b-1) \nonumber \\
  & =  2a^3+2b^3-3(a^2+b^2)+a+b-(a+b)(a+b-1)(2a+2b-1) \nonumber \\
  & =  2(a+b)(a^2-ab+b^2)-3(a+b)^2+6ab+a+b-(a+b)(a+b-1)(2a+2b-1) \nonumber \\
  & =  (a+b)(2a^2-2ab+2b^2-3a-3b+1)+6ab-(a+b)[2(a+b)^2-3(a+b)+1] \nonumber \\
  & =  (a+b)[2a^2-2ab+2b^2-2(a+b)^2]+6ab \nonumber \\
  & =  (a+b)(-6ab)+6ab \nonumber \\
  & =  -6ab(a+b-1) \nonumber \text{,}
\end{align}
The denominator is
\begin{align}
{\rm denominator}=3(a+b)^2(a+b-1) \text{.}
\end{align}
So, the second term is
\begin{align}
{\rm 2nd\;term}=-\frac{2ab}{(a+b)^2} \text{,}
\end{align}
which is negative. When $L\ge 2$, the denominator is positive, and the numerator takes the maximum value at $L=2$. So
\begin{align}
{\rm maximum\;of\;numerator}= & a(a-1)(2a-1)+b(b-1)(2b-1)-\tilde{L}(\tilde{L}-1)(2\tilde{L}-1) \nonumber \\
  = & a(a-1)(2a-1)+b(b-1)(2b-1)-(a+b+1)(a+b)(2a+2b+1) \nonumber \\
  = & (a+b)(2a^2-2ab+2b^2-3a-3b+1)+6ab-(a+b+1)(a+b)(2a+2b+1) \nonumber \\
  = & (a+b)(2a^2-2ab+2b^2-3a-3b+1)+6ab-(a+b)[2(a+b)^2+3(a+b)+1] \nonumber \\
  = & (a+b)(-6ab-6a-6b)+6ab \nonumber \\
  = & -6(a^2b+a^2+ab+ab^2+ab+b^2-ab) \nonumber \\
  = & -6(a^2b+a^2+ab+ab^2+b^2) \text{.}
\end{align}
which is negative.

\section{Transport of Charged Particles}

\subsection{Langevin-Type Stochastic Process}

The probability ${\cal P}({\bf N},t)$ that the cells contain the particle numbers ${\bf N}$ for time $t$ obeys the master equation
\begin{align}
\frac{{\rm d}{\cal P}}{{\rm d}t}=\sum_{i=0}^L\left[\left({\rm e}^{+\partial_{N_i}}{\rm e}^{-\partial_{N_{i+1}}}-1\right)W_i^{(+)}+\left({\rm e}^{-\partial_{N_i}}{\rm e}^{+\partial_{N_{i+1}}}-1\right)W_i^{(-)}\right]{\cal P} \text{.} \label{eq_master_equation}
\end{align}
When $N_i\gg 1$, the operators $\exp(\pm\partial_{N_i})$ can be expanded up to the second order in the partial derivatives $\partial_{N_i}$ in Eq.~(\ref{eq_master_equation}). In this way, we get the Fokker-Planck equation
\begin{align}
\partial_t{\mathscr P}=\sum_{i=0}^L\Bigg\{ & -\partial_{N_i}\left[\left(W_{i-1}^{(+)}-W_{i-1}^{(-)}-W_i^{(+)}+W_i^{(-)}\right){\mathscr P}\right] \nonumber \\
& +\partial_{N_i}^2\left[\frac{1}{2}\left(W_{i-1}^{(+)}+W_{i-1}^{(-)}+W_i^{(+)}+W_i^{(-)}\right){\mathscr P}\right]+\partial_{N_i}\partial_{N_{i+1}}\left[-\left(W_i^{(+)}+W_i^{(-)}\right){\mathscr P} \right]\Bigg\} \text{.}
\end{align}
for the time evolution of the probability density ${\mathscr P}$. This shows that the variables $N_i$ obeys the following stochastic differential equations of Langevin type,
\begin{align}
\frac{{\rm d}N_i}{{\rm d}t}=F_{i-1}-F_i \text{,}
\end{align}
expressed in terms of the fluxes
\begin{align}
F_i=W_i^{(+)}-W_i^{(-)}+\sqrt{W_i^{(+)}+W_i^{(-)}}\xi_i(t) \text{,}
\end{align}
and the Gaussian white noises $\xi_i(t)$ satisfying the properties:
\begin{align}
& \langle\xi_i(t)\rangle=0 \text{,} \\
& \langle\xi_i(t)\xi_j(t')\rangle=\delta_{ij}\delta(t-t') \text{.}
\end{align}

\subsection{The Case of Low-Density Limit}

\par In the low-density limit, the fluctuating electric field is approximated as a static background field. In this case, the energy difference associated with particle transitions between discretized cells are given by
\begin{align}
& \Delta U_{i,i+1}=e(\Phi_{i+1}-\Phi_i)=-\frac{e(\Phi_{\rm L}-\Phi_{\rm R})}{L+1} \text{,} \\
& \Delta U_{i+1,i}=e(\Phi_i-\Phi_{i+1})=+\frac{e(\Phi_{\rm L}-\Phi_{\rm R})}{L+1} \text{,}
\end{align}
which can be uniformly expressed as
\begin{align}
\Delta U^{(\pm)}=\mp\frac{e(\Phi_{\rm L}-\Phi_{\rm R})}{L+1} \text{.}
\end{align}
Correspondingly, $\psi(\Delta U)$ is denoted as
\begin{align}
\psi^{(\pm)}=\psi(\Delta U^{(\pm)})=\frac{\beta\Delta U^{(\pm)}}{\exp\left(\beta\Delta U^{(\pm)}\right)-1} \text{.}
\end{align}
The local transition rates are thus expressed as
\begin{align}
& W_i^{(+)}=\frac{D}{\Delta x^2}\psi^{(+)}N_i=k_+N_i \text{,} \\
& W_i^{(-)}=\frac{D}{\Delta x^2}\psi^{(-)}N_{i+1}=k_-N_{i+1} \text{,}
\end{align}
with the rate constants $k_+$ and $k_-$ defined as obvious. The rates corresponding to the transitions from the reservoirs to the system are 
\begin{align}
W_0^{(+)}=k_+\bar{N}_{\rm L} \text{,}\hspace{1cm} W_L^{(-)}=k_-\bar{N}_{\rm R} \text{.}
\end{align}
Clearly, all these transition rates are determined locally, and as such the transport system in the low-density limit is linear. We consider the time evolution of the probability
\begin{align}
{\cal P}(Z,N_1,\cdots,N_L,t) 
\end{align}
that the cells contain given particle numbers and that the signed cumulated number $Z$ of particles is transferred from the $I$-th to the $(I+1)$-th cells during time interval $[0,\,t]$. This probability is ruled by the following master equation,
\begin{align}
\frac{{\rm d}{\cal P}}{{\rm d}t}=& k_+\bar{N}_{\rm L}\left({\rm e}^{-\partial_{N_1}}-1\right){\cal P}+k_-\left({\rm e}^{+\partial_{N_1}}-1\right)N_1{\cal P}+k_+\left({\rm e}^{+\partial_{N_1}}{\rm e}^{-\partial_{N_2}}-1\right)N_1{\cal P} \nonumber \\
& +\sum_{i=2}^{I-1}\left[k_-\left({\rm e}^{+\partial_{N_i}}{\rm e}^{-\partial_{N_{i-1}}}-1\right)N_i{\cal P}+k_+\left({\rm e}^{+\partial_{N_i}}{\rm e}^{-\partial_{N_{i+1}}}-1\right)N_i{\cal P}\right] \nonumber \\
& +k_-\left({\rm e}^{+\partial_{N_I}}{\rm e}^{-\partial_{N_{I-1}}}-1\right)N_I{\cal P}+k_+\left({\rm e}^{+\partial_{N_I}}{\rm e}^{-\partial_{N_{I+1}}}{\rm e}^{-\partial_Z}-1\right)N_I{\cal P} \nonumber \\
& +k_-\left({\rm e}^{+\partial_{N_{I+1}}}{\rm e}^{-\partial_{N_I}}{\rm e}^{+\partial_Z}-1\right)N_{I+1}{\cal P}+k_+\left({\rm e}^{+\partial_{N_{I+1}}}{\rm e}^{-\partial_{N_{I+2}}}-1\right)N_{I+1}{\cal P} \nonumber \\
& +\sum_{i=I+2}^{L-1}\left[k_-\left({\rm e}^{+\partial_{N_i}}{\rm e}^{-\partial_{N_{i-1}}}-1\right)N_i{\cal P}+k_+\left({\rm e}^{+\partial_{N_i}}{\rm e}^{-\partial_{N_{i+1}}}-1\right)N_i{\cal P}\right] \nonumber \\
& +k_-\bar{N}_{\rm R}\left({\rm e}^{-\partial_{N_L}}-1\right){\cal P}+k_-\left({\rm e}^{+\partial_{N_L}}{\rm e}^{-\partial_{N_{L-1}}}-1\right)N_L{\cal P}+k_+\left({\rm e}^{+\partial_{N_L}}-1\right)N_L{\cal P} \text{.}
\end{align}
Using the method of Ref.~\cite{Gaspard_JStatMech_2018a}, this master equation is solved by introducing the moment generating function
\begin{align}
G(\eta,s_1,\cdots,s_L,t)=\sum_{Z,N_1,\cdots,N_L}\eta^Z\prod_is_i^{N_i}{\cal P}(Z,N_1,\cdots,N_L,t) \text{.}
\end{align}
where
\begin{align}
\eta={\rm e}^{-\lambda} \text{,}
\end{align}
and $\lambda$ is the counting parameter for the particle transfers $Z$. This moment generating function is ruled by the following first-order partial differential equation,
\begin{align}
\partial_tG & + \left[k_-(s_1-1)+k_+(s_1-s_2)\right]\partial_{s_1}G \nonumber \\
& +\sum_{i=2}^{I-1}\left[k_-(s_i-s_{i-1})+k_+(s_i-s_{i+1})\right]\partial_{s_i}G \nonumber \\
& +\left[k_-(s_I-s_{I-1})+k_+(s_I-\eta s_{I+1})\right]\partial_{s_I}G \nonumber \\
& +\left[k_-(s_{I+1}-\eta^{-1}s_I)+k_+(s_{I+1}-s_{I+2})\right]\partial_{s_{I+1}}G \nonumber \\
& +\sum_{i=I+1}^{L-1}\left[k_-(s_i-s_{i-1})+k_+(s_i-s_{i+1})\right]\partial_{s_i}G \nonumber \\
& + \left[k_-(s_L-s_{L-1})+k_+(s_L-1)\right]\partial_{s_L}G \nonumber \\
& =\left[k_+\bar{N}_{\rm L}(s_1-1)+k_-\bar{N}_{\rm R}(s_L-1)\right]G \text{,}
\end{align}
which, in vectoral notations, can be written in the following form,
\begin{align}
\partial_tG+({\boldsymbol{\sf L}}\cdot{\bf s}+{\bf f})\cdot\partial_{\bf s}G=({\bf g}\cdot{\bf s}+h)G \label{eq_differential_equation}
\end{align}
where
\begin{align}
{\boldsymbol{\sf L}}\equiv
\begin{pmatrix}
k_-+k_+ & -k_+ & & & & & & \\
-k_- & k_-+k_+ & -k_+ & & & & & \\
& \ddots & \ddots & \ddots & & & & \\
& & -k_- & k_-+k_+ & -\eta k_+ & & & \\
& & & -\eta^{-1}k_- & k_-+k_+ & -k_+ & & \\
& & & & \ddots & \ddots & \ddots & \\
& & & & & -k_- & k_-+k_+ & -k_+ \\
& & & & & & -k_- & k_-+k_+
\end{pmatrix} \text{,}
\end{align}
\begin{align}
{\bf s}\equiv
\begin{pmatrix}
s_1 \\
s_2 \\
\vdots \\
s_{L-1} \\
s_L
\end{pmatrix}\text{,}
\hspace{1cm}
{\bf f}\equiv-
\begin{pmatrix}
k_- \\
0 \\
\vdots \\
0 \\
k_+
\end{pmatrix}\text{,}
\hspace{2cm}
{\bf g}\equiv
\begin{pmatrix}
k_+\bar{N}_{\rm L} \\
0 \\
\vdots \\
0 \\
k_-\bar{N}_{\rm R}
\end{pmatrix}\text{,}
\end{align}
and
\begin{align}
h\equiv-k_+\bar{N}_{\rm L}-k_-\bar{N}_{\rm R} \text{.}
\end{align}
The parameter $\eta$ in matrix ${\boldsymbol{\sf L}}$ appears in the $I$-th and the $(I+1)$-th rows. From the matrix ${\boldsymbol{\sf L}}$, we can define ${\boldsymbol{\sf L}}_0$ by setting $\eta=1$ and thus $\lambda=0$. So, we have the relations
\begin{align}
& {\bf f}=-{\boldsymbol{\sf L}}_0\cdot{\bf 1} \text{,} \\
& h=-{\bf g}\cdot{\bf 1} \text{,}
\end{align}
where ${\bf 1}$ denotes the vector with all entries equal to one. Besides, the stationary values of particle numbers are given by
\begin{align}
{\bf\Gamma}_0={\boldsymbol{\sf L}}_0^{-1{\rm T}}\cdot{\bf g} \text{.} \label{eq_Gamma_0}
\end{align}
The first-order partial differential equation~(\ref{eq_differential_equation}) can be solved by the method of characteristics. The equations for the characteristics are given by
\begin{align}
& \frac{{\rm d}{\bf s}}{{\rm d}t}={\boldsymbol{\sf L}}\cdot{\bf s}+{\bf f} \text{,} \label{eq_characteristics_equation} \\
& \frac{{\rm d}G}{{\rm d}t}=({\bf g}\cdot{\bf s}+h)G \text{.} \label{eq_G_equation}
\end{align}
The solution of Eq.~(\ref{eq_characteristics_equation}) gives the characteristics
\begin{align}
{\bf s}={\rm e}^{{\boldsymbol{\sf L}}t}\cdot\left[{\bf s}_0+{\boldsymbol{\sf L}}^{-1}\cdot\left({\boldsymbol{\sf I}}-{\rm e}^{-{\boldsymbol{\sf L}}t}\right)\cdot{\bf f}\right] \text{.}
\end{align}
Replacing in Eq.~(\ref{eq_G_equation}), we obtain after integration that
\begin{align}
G=G_0\exp\left[{\bf g}\cdot{\boldsymbol{\sf L}}^{-1}\cdot\left({\boldsymbol{\sf I}}-{\rm e}^{-{\boldsymbol{\sf L}}t}\right)\cdot\left({\bf s}+{\boldsymbol{\sf L}}^{-1}\cdot{\bf f}\right)+\left(h-{\bf g}\cdot{\boldsymbol{\sf L}}^{-1}\cdot{\bf f}\right)t\right] \text{,}
\end{align}
where ${\boldsymbol{\sf I}}$ denotes the identity matrix. The initial condition being the Poisson distribution describing the steady state and the counter reset to zero $Z=0$, we have that
\begin{align}
G_0(\eta,{\bf s}_0)={\rm e}^{{\bf\Gamma}_0\cdot({\bf s}_0-{\bf 1})} \text{.}
\end{align}
The solution of Eq.~(\ref{eq_differential_equation}) is thus given by
\begin{align}
G(\eta,{\bf s},t)= & \exp\left[{\bf g}\cdot{\boldsymbol{\sf L}}^{-1}\cdot\left({\boldsymbol{\sf I}}-{\rm e}^{-{\boldsymbol{\sf L}}t}\right)\cdot\left({\bf s}+{\boldsymbol{\sf L}}^{-1}\cdot{\bf f}\right)+\left(h-{\bf g}\cdot{\boldsymbol{\sf L}}^{-1}\cdot{\bf f}\right)t\right] \nonumber \\
&\times\exp\Big\{{\bf\Gamma}_0\cdot\left[{\rm e}^{-{\boldsymbol{\sf L}}t}\cdot{\bf s}-{\boldsymbol{\sf L}}^{-1}\cdot\left({\boldsymbol{\sf I}}-{\rm e}^{-{\boldsymbol{\sf L}}t}\right)\cdot{\bf f}-1\right]\Big\} \text{.} \label{eq_G}
\end{align}
The cumulant generating function of the signed cumulated transfers of particles from the $I$-th to the $(I+1)$-th cell is defined as
\begin{align}
Q(\lambda)\equiv\lim_{t\to\infty}-\frac{1}{t}\ln\left[G(\eta={\rm e}^{-\lambda},{\bf 1},t)\right]={\bf g}\cdot\left({\bf 1}+{\boldsymbol{\sf L}}^{-1}\cdot{\bf f}\right) \text{,} \label{eq_generating_function}
\end{align}
where the positivity of ${\boldsymbol{\sf L}}_0$ has been used to obtain the explicit expression. We observe that
\begin{align}
{\boldsymbol{\sf L}}={\bf M}\cdot{\boldsymbol{\sf L}}_0\cdot{\bf M}^{-1} \text{,}
\end{align}
where
\begin{align}
{\bf M}=\eta{\bf P}_{\rm L}+{\bf P}_{\rm R}
\end{align}
with the projection matrices
\begin{align}
{\bf P}_{\rm L}=
\begin{pmatrix}
1 & \cdots & 0 & 0 & \cdots & 0 \\
\vdots & \ddots & \vdots & \vdots & \ddots & \vdots \\
0 & \cdots & 1 & 0 & \cdots & 0 \\
0 & \cdots & 0 & 0 & \cdots & 0 \\
\vdots & \ddots & \vdots & \vdots & \ddots & \vdots \\
0 & \cdots & 0 & 0 & \cdots & 0 \\
\end{pmatrix}
\hspace{1cm}\text{and}\hspace{1cm}
{\bf P}_{\rm R}=
\begin{pmatrix}
0 & \cdots & 0 & 0 & \cdots & 0 \\
\vdots & \ddots & \vdots & \vdots & \ddots & \vdots \\
0 & \cdots & 0 & 0 & \cdots & 0 \\
0 & \cdots & 0 & 1 & \cdots & 0 \\
\vdots & \ddots & \vdots & \vdots & \ddots & \vdots \\
0 & \cdots & 0 & 0 & \cdots & 1 \\
\end{pmatrix} \text{.}
\end{align}
The identity matrix in ${\bf P}_{\rm L}$ is of dimension $I\times I$, while the identity matrix in ${\bf P}_{\rm R}$ is $(L-I)\times(L-I)$. Since the projection matrices satisfy the condition ${\bf P}_{\rm L}+{\bf P}_{\rm R}={\boldsymbol{\sf I}}$, we thus have
\begin{align}
& {\bf M}={\boldsymbol{\sf I}}+(\eta-1){\bf P}_{\rm L} \text{,} \label{eq_M} \\
& {\bf M}^{-1}={\boldsymbol{\sf I}}+(\eta^{-1}-1){\bf P}_{\rm L} \text{.} \label{eq_M_inverse}
\end{align}
From the above related expressions, the cumulant generating function~(\ref{eq_generating_function}) can be written in the following form,
\begin{align}
Q(\lambda)={\bf g}\cdot\left[{\boldsymbol{\sf I}}-{\bf M}\cdot{\boldsymbol{\sf L}}_0^{-1}\cdot{\bf M}^{-1}\cdot{\boldsymbol{\sf L}}_0 \right]\cdot{\bf 1} \text{.}
\end{align}
Because of Eqs.~(\ref{eq_M})-(\ref{eq_M_inverse}), we find that
\begin{align}
Q(\lambda,t)={\bf g}\cdot\left[(1-\eta){\bf P}_{\rm L}+(1-\eta^{-1}){\boldsymbol{\sf L}}_0^{-1}\cdot{\bf P}_{\rm L}\cdot{\boldsymbol{\sf L}}_0-(2-\eta-\eta^{-1}){\bf P}_{\rm L}\cdot{\boldsymbol{\sf L}}_0^{-1}\cdot{\bf P}_{\rm L}\cdot{\boldsymbol{\sf L}}_0 \right]\cdot{\bf 1} \text{.}
\end{align}
Using Eq.~(\ref{eq_Gamma_0}) and ${\bf P}_{\rm R}={\boldsymbol{\sf I}}-{\bf P}_{\rm L}$, the cumulant generating function becomes
\begin{align}
Q(\lambda)=W_+\left(1-{\rm e}^{-\lambda}\right)+W_-\left(1-{\rm e}^{+\lambda}\right) \text{,} \label{eq_Q2}
\end{align}
with the global transition rates given by
\begin{align}
& W_+={\bf\Gamma}_0\cdot{\boldsymbol{\sf L}}_0\cdot{\bf P}_{\rm L}\cdot{\boldsymbol{\sf L}}_0^{-1}\cdot{\bf P}_{\rm R}\cdot{\boldsymbol{\sf L}}_0\cdot{\bf 1} \text{,} \label{eq_W_plus} \\
& W_-={\bf\Gamma}_0\cdot{\boldsymbol{\sf L}}_0\cdot{\bf P}_{\rm R}\cdot{\boldsymbol{\sf L}}_0^{-1}\cdot{\bf P}_{\rm L}\cdot{\boldsymbol{\sf L}}_0\cdot{\bf 1} \label{eq_W_minus} \text{.}
\end{align}
The cumulant generating function~(\ref{eq_Q2}) has exactly the same form as Eq.~(\ref{eq_Q1}). This indicates that the long-time behavior of the particle transport can be captured by the coarse-grained model with two equivalent global transition rates $W_+$ and $W_-$. These two global transition rates~(\ref{eq_W_plus})-(\ref{eq_W_minus}) can be developed as
\begin{align}
W_+=k_+^2\bar{N}_{\rm L}\left({\boldsymbol{\sf L}}_0^{-1}\right)_{1L} \hspace{1cm}\text{and}\hspace{1cm} W_-=k_-^2\bar{N}_{\rm R}\left({\boldsymbol{\sf L}}_0^{-1}\right)_{L1} \text{.}
\end{align}
Inverting the matrix ${\boldsymbol{\sf L}}_0$, we get
\begin{align}
\left({\boldsymbol{\sf L}}_0^{-1}\right)_{ij}=
\begin{cases}
\frac{k_+^{j-i}\left(k_+^i-k_-^i\right)\left(k_+^{L+1-j}-k_-^{L+1-j}\right)}{\left(k_+-k_-\right)\left(k_+^{L+1}-k_-^{L+1}\right)} & \text{if}  \quad i\leq j \text{,} \\
& \vspace{-0.3cm} \\
\frac{k_-^{i-j}\left(k_+^j-k_-^j\right)\left(k_+^{L+1-i}-k_-^{L+1-i}\right)}{\left(k_+-k_-\right)\left(k_+^{L+1}-k_-^{L+1}\right)} & \text{if}  \quad i>j \text{.}
\end{cases}
\label{eq_inverse_C}
\end{align}
So, the two global transition rates are calculated as
\begin{align}
& W_+=\bar{N}_{\rm L}\frac{k_+^{L+1}(k_+-k_-)}{k_+^{L+1}-k_-^{L+1}}=\frac{D\bar{N}_{\rm L}}{\Delta x^2(L+1)}\frac{\beta e(\Phi_{\rm L}-\Phi_{\rm R})}{1-\exp\left[-\beta e(\Phi_{\rm L}-\Phi_{\rm R})\right]} \text{,} \\
& W_-=\bar{N}_{\rm R}\frac{k_-^{L+1}(k_+-k_-)}{k_+^{L+1}-k_-^{L+1}}=\frac{D\bar{N}_{\rm R}}{\Delta x^2(L+1)}\frac{\beta e(\Phi_{\rm R}-\Phi_{\rm L})}{1-\exp\left[-\beta e(\Phi_{\rm R}-\Phi_{\rm L})\right]} \text{.}
\end{align}

\printbibliography[title={References}]

\end{document}